\documentclass[acmsmall]{acmart}
\usepackage{csquotes}
\usepackage{makecell}
\usepackage{enumitem}
\usepackage{multirow,siunitx}
\DeclareUnicodeCharacter{0306}{\u{}}
\AtBeginDocument{%
  }

\setcopyright{cc}
\setcctype{by}
\acmJournal{PACMHCI}
\acmYear{2026} \acmVolume{10} \acmNumber{6} \acmArticle{CSCW133}
\acmMonth{10} \acmDOI{10.1145/3816981}

\received{May 13, 2025}
\received[revised]{January 13, 2026}
\received[accepted]{March 17, 2026}

\newcommand{\edit}[1]{\textcolor{black}{#1}}

\begin{document}

\title[Understanding Community, Challenges, and Consequences in Digitally-facilitated Labor Organizing]{Organizing in the Digital Age: Understanding Community, Challenges, and Consequences in Digitally-facilitated Labor Organizing}


\author{Frederick Reiber}
\email{freddyr@bu.edu}
\affiliation{%
  \institution{Boston University}
  \city{Boston}
  \state{Massachusetts}
  \country{USA}
}

\author{Alishah Chator}
\affiliation{%
  \institution{Baruch College}
  \city{New York}
  \state{New York}
  \country{USA}
}
\author{Dana Calacci}
\affiliation{%
    \institution{Penn State University}
    \city{University Park}
    \state{Pennsylvania}
    \country{USA}
}

\author{Allison McDonald}
\affiliation{%
  \institution{Boston University}
  \city{Boston}
  \state{Massachusetts}
  \country{USA}
}


\begin{abstract}
    The contemporary American labor force is highly dispersed, necessitating the use of digital communication tools to bridge spatial and temporal gaps in union organizing. This study provides an in-depth analysis of how workers within various labor unions utilize digital, text-based communication platforms---including Discord, WhatsApp, and Slack---for labor organizing. Through 17 qualitative interviews, we examine the challenges and opportunities presented by digital organizing, identifying both technical and social obstacles. Our findings reveal that although digital tools are integral to contemporary labor successes, they also introduce new complexities, such as navigating technical security, managing information overload, and building trust and consensus. Based on these insights, we draw connections to broader understandings of digital organizing and the role of digital tools in unions. 
\end{abstract}



\maketitle

\newcommand{\asc}[1]{\textcolor{magenta}{ASC: #1}}
\newcommand{\amc}[1]{\textcolor{purple}{\textbf{AMc:} #1}}
\newcommand\todo[1]{\textbf{\textcolor{orange}{[[#1]]}}}

\section{Introduction}
\label{sec:intro}

On August 23, 2021 Starbucks baristas in Buffalo, New York posted a letter addressed to then-Starbucks President and CEO Kevin Johnson. Its first sentence read, ``we believe that there can be no true partnership without power-sharing and accountability,'' with the rest of the letter detailing the newly formed organizing committee's desire to unionize the Buffalo Starbucks partners. Workers would indeed unionize later that year, winning their vote with 19 to 8 in favor on December 9, 2021~\cite{canella_networked_2023}. Public coverage of the vote then set off a powder keg of worker frustration at Starbucks, with thousands of workers and hundreds of stores signing petitions to unionize following Buffalo's vote. For workers at Starbucks, who are relatively spread out and lack a central workplace, digital tools have been and continue to be critical to their union drive, facilitating much of their discussion, learning, and organizing.

In his new book \textit{We Are The Union}, labor studies professor Eric Blanc highlights how workers are revitalizing the American labor movement through the use of digital technologies~\cite{blanc_we_2025}. 
Unlike in the 1930s---labor's strongest point historically---historical community-driven approaches to organizing are no longer viable. 
Workers and workplaces are far too spread out, lacking the collective space that was critical to labor's original rise. To combat these spatial and temporal challenges, workers have instead turned to digital tools. Many of the current organizing campaigns within the United States heavily utilize digital technologies like WhatsApp, Slack, or Zoom to help facilitate their organizing. To quote a local Boston labor organizer, ``\textit{There is no organizing without WhatsApp.}'' 

This paper presents an in-depth study of the experiences, challenges, and outcomes of \textit{workers} using digital text-based communication tools, including Discord and WhatsApp, for the purposes of organizing within a formal organized labor context. Although previous literature has highlighted the complex and often troubled relationship between technology and labor, with technology often used to curb organized labor~\cite{hennebert_what_2021, parker_adapt_2022, cross_stopping_2024, reiber_unionBusting_2025} or illustrating the importance of technology in labor organizing~\cite{asad_illegitimate_2015, blanc_we_2025}, there is little work that centers on workers and their interactions with the technological tools used for labor organizing. This paper fills that gap, focusing on how workers within a number of labor unions are using digital text-based tools like Discord, WhatsApp, and Slack to facilitate labor union organizing. In doing so, we contribute to a number of active research threads in both CSCW and labor relations on the role technology plays in organized labor~\cite{nissim_future_2021, schradie_context_2021, khovanskaya_cyberunion_2023}. To build this knowledge, we conducted 17 in-depth, semi-structured interviews with workers based in the United States across industries, including tech, higher education, and legal services, spanning varying levels of union participation.

\subsection{Research Questions}

Building on recent literature in computing and organized labor, we analyze the production of outcomes as a dialectical process between organization and technology, recognizing that a labor union’s identity and structure come into conflict with, and are shaped by, technology~\cite{kellogg_algorithms_2020, geelan_introduction_2021}. We see this interaction as a complicated process, which is shaped by more than just the technical features of the technology or the structural elements of the union. This is the basis for our first research question, which aims to explore how this interaction operates in practice and the outcomes of that interaction. To help structure this investigation, we analyze three different components of how labor unions interact with digital communication tools: how labor unions \textit{structure} their digital spaces, the \textit{dynamics} of these digital spaces, and the broader \textit{cultural} shifts workers associate with digital organizing.

Additionally, we investigate how digital communication impacts tensions commonly found in unions between rank-and-file members and ``union bureaucracy.'' Numerous labor activists and academics have recognized that current union officials tend to favor working with employers or governments over prioritizing workers~\cite{mcalevey_no_2016, mcilroy_marxism_2014}. Conversely, rank-and-file members tend to be more combative towards management and favor more direct forms of action~\cite{camfield_what_2013, bash_letter_2024, brown_uaw_2024}. We examine how the use of a digital communication platform for inter-member communication shifts these dynamics between union professionals and rank-and-file members.

Formally, our research questions are:

\begin{itemize}
    \item \textbf{RQ1}: From a rank-and-file worker's perspective, what does the outcome of labor organizing through digital information and communication technologies look like? Specifically, we focus on:
    \begin{itemize}
        \item[$\blacksquare$] Structure\textemdash how are workers and unions constructing these digital spaces?
        \item[$\blacksquare$] Dynamics\textemdash what are the patterns of interaction, behaviors, and relationships between individuals within digital union spaces? 
        \item[$\blacksquare$] Culture\textemdash how do digital spaces affect the perception of solidarity, leadership, and participation that shape regular union actions and responses?
    \end{itemize}
    \item \textbf{RQ2}: How does the use of digital communication tools for organizing affect the power dynamics between different stakeholders within unions, specifically between rank-and-file members and professional full-time union staff?
\end{itemize}

\subsection{Contributions}
This research is of interest to a number of groups, including both the CSCW community and the labor organizing community. For the CSCW community, we provide detailed insights into how activist and labor communities adapt chat platforms to fit their needs and the challenges associated with organizing through currently-available digital tools. This analysis also responds to calls for work on collective action, organized labor, and technology~\cite{sum_its_2024, hennebert_what_2021}. For the labor community, we provide a detailed analysis of the successes and challenges of organizing digitally. This enables labor organizers to make more informed decisions about how to structure, use, and organize within digital spaces, as well as add new techniques to organizing efforts within their communities.

We also believe that this research could not come at a more critical time. Interest in labor unions is on the rise, with union support at its highest level in decades~\cite{inc_us_2022}. This is not just a theoretical rise, as nearly 500,000 workers have unionized in the past two years~\cite{rhinehart_new_2024}. Additionally, labor activists have started digital labor projects like the Emergency Workplace Organization Council (EWOC)~\footnote{https://workerorganizing.org/}. This group of labor activists, scholars, and organizers helps provide everyday workers with tools and training to organize their workplace~\cite{blanc_emergency_2024} through a large networked Slack. As labor organizers continue to overcome the challenges of organizing in the highly spread out workforce, having an in-depth understanding of how to best use and design digital tools for organizing is critical to union success.

\section{Related Work}
\label{sec:related}

This paper draws on and expands two threads of research central to the CSCW community: designing and understanding digital tools for workplace organizing and the effects of technology on the workplace. 

\subsection{\edit{Digital Tools for Organizing}}

Scholars in both the CSCW and CHI communities have had significant interest in designing and understanding tools for organizing~\cite{calacci_organizing_2022, ghoshal_design_2019, ghoshal_toward_2020}. One thread of this work focuses on providing spaces for organizing in the gig economy. Gig workers are often incredibly dispersed, making organizing and collective action difficult. Research projects like Turkopticon~\cite{irani_turkopticon_2013} and Dynamo~\cite{salehi_we_2015} provide platform workers with a space to share information and build collective wisdom, trying to correct the power imbalance coded into tech-facilitated work. More recent projects have centered on providing data tools for organizers in the workforce. For instance, tools like FairFare~\cite{calacci_fairfare_2025} or the Shipt Calculator~\cite{calacci_bargaining_2022} provide organizers the tools to audit the obtuse algorithmic systems of gig work. These tools, developed in collaboration with worker advocacy groups, provide workers and organizers both data and analytics, helping to correct data imbalances within labor organizing. 

Scholars have also investigated how organizers use general-purpose tools for organizing. For example, Bonini et al.~\cite{bonini_cooperative_2024} investigated the use of instant messaging apps among food delivery gig workers, finding that the tools helped workers develop solidarity and facilitate resistance. Thuppilikkat, Dhar, and Chandra~\cite{thuppilikkat_union_2024} investigated the role technology played in union struggles within the Kolkata, India taxi-cab service, \edit{highlighting the new models of organizing formed through tech adaptation.} Both of these works also stress the importance of digital communication tools in overcoming the spatial issues of gig work. In Asad and Dantec's work on ``Illegitimate Civic Participation,'' they highlight the role of digital communication technologies in housing activist groups acting outside of standard democratic processes~\cite{asad_illegitimate_2015}. Other work has looked at the role data plays in community organizing in immigration and reproductive justice~\cite{pei_for_2024} or the usage of Twitter bots to activate potential volunteers~\cite{savage_botivist_2016}. Recent scholarship has also investigated the design \textit{conflicts} between grassroots organizing and commercial technologies, finding that they clash with the practices and values of organizing communities. For example, Ghoshal et al.~\cite{ghoshal_role_2019} work towards developing a theory of grassroots technological practice, highlighting that the organizing goals of inclusivity clash with the inherent exclusivity of digital tools, and Nguyen et al.~\cite{nguyen_it_2025} investigate the technological practices of mutual aid networks, finding challenges in preserving community understanding when adopting technologies into their work.

While informative for our work, these studies look at significantly less formal modes of community organizing. Labor unions in the United States---unlike immigration or housing activists---are significantly more regulated, with federal law dictating what unions can and cannot do~\cite{bennett_what_2007}. Unions also have a significant bureaucratic and organizational structure that presents a new set of significantly understudied challenges for activist and tech interaction~\cite{bennett_what_2007}. This positions us closer to the small collection of CSCW and CHI studies that have investigated the relationship between more formalized types of collective action and computing. For example, Khovanskaya, Senger, and Dombrowski's 2020 CHI paper, ``Bottom-Up Organizing with Tools from On High''~\cite{khovanskaya_bottom-up_2020}, looked into the data practices of labor organizers, finding conflicts between union organizing and the hierarchical nature of unions and their data tools. Kapoor et al.~\cite{kapoor_weaving_2022} explored the privacy practices of tech labor organizers, finding numerous technical challenges, both in terms of individual and group privacy, while Spektor et al.~\cite{spektor_data_2025} investigated the role of data within more formal structures of organizing. These threads help inform our work by providing insights into how labor unions use, and the challenges that come with, digital data tools. Our work builds upon this by focusing in on digital communication tools---another part of the digital organizing equation.

Finally, we also draw and expand upon a number of labor relations papers. Researchers in sociology and labor relations have explored the idea of using digital technologies to help increase union participation~\cite{shostak_cyberunion_2015}, highlighting the ability for digital technologies to increase union democracy~\cite{greene_commentary_2003} by providing easy access to participation~\cite{greene_possibilities_2003} or increasing transparency~\cite{diamond_will_2002}. More recent lines of work have also investigated the use of social media as a tool for union engagement~\cite{thornthwaite_unions_2018, panagiotopoulos_social_2015, hodder_unions_2020} and highlight the importance of using localized knowledge. Critically, both of these lines of work have received critique~\cite{hennebert_what_2021} as empirical work has shown digital communication tools to have little effect on participation~\cite{waddington_e-communications_2014} or to even have a stifling effect on debate~\cite{hodder_union_2015}. However, since these papers' publication, the landscape and understanding of union activity has changed significantly, with more calls to center workers in union organizing, necessitating a re-evaluation of these debates~\cite{mcalevey_no_2016, birelma_militant_2023, uetricht_us_2019, blanc_we_2025}.

\subsection{Impacts of Computation on the Workforce}

The CSCW community has a long history of developing and studying technologies within the workplace~\cite{fox_worker-centered_2020, ehn_work-oriented_1988}. Of particular relevance to our work are issues of digital workplace surveillance, which has become ubiquitous with the rise of digital work technologies~\cite{zuboff_surveillance_2019, kellogg_algorithms_2020}. For example, work by Sum, Shi, and Fox~\cite{sum_its_2024} explores how workers try to avoid workplace surveillance, while Chowdhary et al.~\cite{chowdhary_can_2023} question if workers have meaningful consent when dealing with workplace well-being technologies. Other threads of HCI scholarship look into how workers react to technological advances in the workplace. For example, researchers have looked at how ``essential'' workers have reacted to AI-driven shifts in the labor process~\cite{kang_stories_2022} or issues relating to algorithmic management~\cite{zhang_algorithmic_2022}. 

Our work provides new insights into the challenges posed by technology in the workplace and its effects on the labor force. Labor unions continue to play a crucial role in helping workers shape the terms and conditions of their employment, and many unions are actively responding to the changing nature of work brought about by digital technologies~\cite{cross_stopping_2024}. However, as numerous scholars have noted, these same technological developments\textemdash ranging from algorithmic management to data surveillance\textemdash have made it increasingly difficult to organize and sustain labor unions~\cite{garden_labor_2018, bernhardt_data-driven_2023}. Our work helps to fill gaps in the understanding of computation and workplace dynamics by studying how organized labor is operating under an increase of computational technologies in modern working conditions.

\section{Labor Unions and Organizing Background}
\label{sec:unionbkg}

\edit{At a high level, American labor unions are a democratic organization of workers who collectively bargain with their employer to secure better terms of employment, with a focus on wages, benefits, working conditions, and job security~\cite{mcalevey_collective_2020}. Unions have played a crucial role in advocating for the rights and well-being of workers and have a long history of influencing labor laws and workplace policies~\cite{loomis_history_2018}. One of the main ways that American labor unions improve working conditions is through the collective bargaining agreement or CBA, a jointly-negotiated contract that sets workplace and employment conditions for a given period of time. Achieving a CBA also requires employer recognition of the union, which can happen either voluntarily---as in, the employer willingly recognizes the union's right to collectively bargain---or through a National Labor Relations Board election~\cite{ferguson_eyes_2008}. Critically, both of these processes---union recognition and collective bargaining---and some of the internal processes of unions are regulated at both the federal and state level~\cite{goldberg_overview_2000, atleson_law_1994}.}   

\subsection{\edit{Union Organizing and Inter-union Tensions}}

\edit{American labor unions exist within a complex weave of tensions both externally and internally. Central to these tensions is what it means to ``organize'' within and for a union, as union organizing brings high risks. Borrowing from renowned scholar and union organizer Jane McAlevey, this paper conceptualizes union organizing as the building of relationships with the goal of transferring power from a small minority (employers) to a more public majority (employees)~\cite{mcalevey_no_2016}. Unlike other definitions of organizing, our focus here is on the development and building of relationships within the workplace, along with the activation of those relationships for power~\cite{han_prisms_2024, grinthal_power_2011}. This power can then manifest in numerous ways, from wearing pro-union pins, signing petitions, to going on strike. When compared to other forms of movement building like ``mobilizing''---which focuses on activating workers who already agree with workplace organizing---organizing focuses on empowering or recruiting people who may have withdrawn from, or would not naturally participate in, activism. These relationships are also critically important for realizing for worker power, as both organizers and prior research have recognized the importance of interpersonal relationships within strike and direct action participation~\cite{mcalevey_no_2016,mcalevey_collective_2020, akkerman_solidarity_2013}. Analysis of said relationships is also where terms like ``shallow,'' ``service,'' or ``militant'' organizing emerge, reflecting attempts to characterize the relationships produced through union organizing. As an example, service unionism characterizes unions that focus on providing a service to their workers, where as militant unions are those that focusing on direct action like strikes.}
 
\edit{One of the ways this definition creates conflict is through the different roles within unions,} often between rank-and-file, dues-paying members and professional union staffers~\cite{levi_union_2009, vassiley_once_2024}. \edit{Because labor unions are formalized structures---federally regulated and defined---tensions often arise around union structure and democracy~\cite{bennett_what_2007}.} Rank-and-file workers, while the backbone of any labor union, are often disconnected from larger labor union debates around bargaining or actions. Unlike full-time officials, rank-and-file organizers operate as volunteers, forcing them to balance their normal work duties with union activity. Rank-and-file organizers are also often paid substantially less than staffers and risk retaliation at their workplace~\cite{darlington_reappraisal_2012}. 

Labor union officials tend to sit in the middle of capital and labor relations, trying to balance their relationships with both sides. While they are invested in improving conditions for workers, union leaders are also concerned with maintaining the union and avoiding lawsuits or state-mandated fines~\cite{darlington_reappraisal_2012}. These tensions often create conflict in organizing as defined by McAlevey. When unions center full-time union staff in labor organizing, workers are less likely to build meaningful relationships with each other, and thus workers' ability to create power is limited. This can also result in significant differences when it comes to union direct action. Realizing worker power often requires work disruptions like walkouts or strikes, but due the fears of hurting professional relationships, union officials may oppose such actions, leading to wildcat strikes---a strike where workers walk off the job without union support~\cite{elk_wave_2018, blanc_strike_2018}.

\edit{These tensions---between organizing and mobilizing and between rank-and-file and staff---frame much of our research, both in understanding how technology shapes organizing, and as a starting point for understanding power within digital union spaces}. \edit{However, we also want to note that these definitions and understanding are contested. For instance, sociologist Kim Voss has pushed back against the idea of inherent frictions between union leaders and workers, citing cases in which union management worked well with rank-and-file members~\cite{voss_democratic_2010, voss_breaking_2000}. We bring this up as an issue of methodological importance: while this common dynamic did inform our research, we made sure to allow for divergent experiences and make sure to not assume this tension in both interviews and analysis.}

\subsection{Labor Unions and Technology}

Over the past 20 to 30 years, unions have adopted a number of digital practices, using and augmenting their organizing with technology. For instance, unions can use complex union management platforms like UnionWare to manage dues and membership information. Other times unions will use tech-blast or voting software to help bring workers into the union and mobilize them for direct actions~\cite{hennebert_what_2021}. \edit{Recognizing the potential power of digital tools, labor scholars have theorized the concept of \textit{digital renewal}: the idea that digital technology could be used to help facilitate union renewal~\cite{waddington_e-communications_2014, carneiro_digital_2022}.} 

Unfortunately, many of these attempts have---according to the scholarly work---failed~\cite{hennebert_what_2021}. In 2008, one of the UK's largest unions, UNISON, attempted to develop and deploy union websites titled ``virtual branches.'' UNISON's locals put significant amounts of work into bringing more workers into organizing through theses digital space. However, as authors Kerr and Waddington find, the development of these tools didn't meaningfully change the landscape of union participation, with some of the locals leaving the project due to its costs~\cite{waddington_e-communications_2014}. As academics and labor organizers ourselves, we see this as a consequence of the technology being driven from the top rather than from the bottom, reflecting a broader pattern in UNISON's history of being less directly engaged with bottom-up, member-led organizing~\cite{waddington_transforming_2009}. Other inquires have also found similar results~\cite{martinez_lucio_new_2003, martinez_lucio_networked_2005}. For instance, in a study on using digital technologies to build solidarity across national borders, the authors found that the use of information and communication technologies intersected with organizational power hierarchies, tensions within communities of practice, and the historical relationships between different groups~\cite{lucio_making_2009}. More plainly put, digital technologies have not revived a labor movement. Thus, if we want technology to aid in union-renewal, we need a better understanding of the dynamics at the bottom---the workers.

There are also a number of tensions within unions that make introducing and studying technology a challenge. Labor unions, especially in the United States, have a very tense relationship with technology~\cite{chaison_information_2002, nissim_future_2021, parker_adapt_2022}. Numerous labor unions, like the International Typographical Union, have been completely destroyed by technological advancement fundamentally reorganizing work~\cite{parker_adapt_2022}. Labor unions---despite a relative high in approval and support~\cite{inc_us_2022}---are also at an incredibly low level of worker coverage. We argue that this is partial driven by technological innovations that have reshaped the workforce and management~\cite{bernhardt_data-driven_2023, kellogg_algorithms_2020}. For example, the gig economy and its theorized ``future of work'' is, as Niels van Doorn puts, a tool ``weaponised... to break the power of organised labour'' by repositioning workers not as employees, but as small business owners engaging in the ``free marketplace''~\cite{doorn_at_2020}. These conflicts between technology and labor also have a long history, as workers dating back to the 19th century have fought against the blind introduction of workplace technology~\cite{linton_luddites_1992} for fear of it reducing their collective power. Because of this history, many union members and organizers, including some of our participants, identify as a Luddites, seeing technology as tool designed to hurt, not help, workers. Therefore, when attempting to introduce, design, or study technology in a labor organizing space, one needs to be critically aware of the tension between these two forces.

\section{Methods}
\label{sec:methods}

To develop this analysis, we conducted 17 semi-structured interviews with union members. Interviews took place on Zoom from January 2025 to March 2025. In interviews, we started with some background questions to help contextualize the unions activities of our participants. We then transitioned to discussions around the use of digital hubs, focusing on the structure, usage, and dynamics of the tools. Finally, we closed with higher-level discussions around the role of these tools in the union's culture and a reflection on digital organizing. Participants were compensated with a $\$$25 prepaid Visa gift card. This study was reviewed and approved by our university's Institutional Review Board (IRB). 

\begin{table}
\small
\setlength\extrarowheight{3pt}
\setlength\tabcolsep{0pt}
\begin{tabular*}{0.95\textwidth}{ @{\extracolsep{\fill}} 
      lcl cc cc @{} }
\hline
\hline
\textbf{Union} & \textbf{Participants} & \textbf{Union Role} & \textbf{Industry} \\
\hline
\multirow{4}{*}{Union 1 }  
    & P1 & Elected officer & \multirow{4}{*}{Higher Education} \\
\cline{2-3}
    & P2 & Active organizer \\
\cline{2-3}
    & P3 & Steward \\
\cline{2-3}
    & P4 & General member \\
\cline{1-4}
\multirow{4}{*}{Union 2 }
    & P6 & General member & \multirow{4}{*}{Technology}\\
\cline{2-3}
    & P7 & Active organizer, Bargaining committee \\
\cline{2-3}
    & P12 & Active organizer \\
\cline{2-3}
    & P13 & Active organizer \\
\hline
\multirow{4}{*}{Union 3} 
    & P8 & Bargaining committee & \multirow{4}{*}{Legal Services} \\
\cline{2-3}
    & P9 & Steward \\
\cline{2-3}
    & P10 & General member \\
\cline{2-3}
    & P16 & General member \\
\hline
\multirow{3}{*}{Union 4}
    & P11 & Bargaining committee & \multirow{3}{*}{Food and Entertainment} \\
\cline{2-3}
    & P14 & General member \\
\cline{2-3}
    & P15 & General member \\
\hline
Union 5 & P17 & Active organizer, Staffer & \multirow{2}{*}{Union Staff}\\
\cline{1-3}
Union 6 & P5 & General member, Staffer \\
\hline
\hline
\end{tabular*}
\caption{Overview of unions and associated participants with participant and union codes, role within the union, and industry}
\Description[A summary of unions and participants]{A table highlighting the roles, membership, and industry of each participant within our study}
\label{table:interviews}
\end{table}

\subsection{Participant Recruitment}

When recruiting, our main criteria were current or recent labor union members, specifically workers \textit{within} unions who had experience participating in union activities through the use of a digital text-based platform. \edit{We define labor union members as workers who did not earn their primary income through elected or leadership positions within a labor union, but whose conditions of employment were either set by a collective bargaining agreement or were in the process of being organized toward one.} \edit{Note that union staffers are still in-scope as, while they work for a union, they can also engage in collective bargaining with the union as their employer.} Due to complicated and often contested definitions of which workers are considered ``in-unit,'' we did not require participants to be active dues-paying members. Instead, we required participants to have participated---on the side of the union---in some sort of union related activity through digital text communication channels. Our definition of union activities was similarly broad. This could range from significant amounts of organizing work, including participation in bargaining or organizing pickets, to simply participating in a few conversations around what unionization or collective bargaining would mean for the workplace.

As one of our research questions deals with union democracy, we aimed to speak to multiple workers within each of the unions we recruited from. We also recruited members who filled different roles within the union to help ensure our data represented multiple perspectives. Additionally, we recruited across a number of different unions, with the goal of getting \edit{a broad set of experiences}. Table~\ref{table:interviews} provides an overview of the unions and associated participants in our study. \edit{We also provide a brief characterization of the workers' roles in the union. ``General member'' refers to union members who participated in union activities but not in any of the planning or organizing of the activities. ``Active organizer'' refers to members who  aided in the planning and organizing of union activities. ``Bargaining committee'' refers the group of people who participated or led bargaining meetings with the employer's representative and ``Elected officer'' or ``Steward'' represent elected worker roles. We also label the two workers employed by a union as ``Staffers'' in addition to their role label within their own union.} Finally, \textit{union 2 is primarily a remote workplace} and is scoped around the local, as they were in the process of restructuring at the time of the interviews. 

Recruitment was done through two main avenues. First, we distributed an advertisement with a QR code in the relevant communication channels of five labor unions via the authors' personal networks. Potential participants would then be taken to a brief Qualtrics form that collected high-level demographic information and an email we could use to schedule an interview. The primary author would then reach out to schedule a time and review our consent protocol. Second, we utilized snowball sampling to expand our access to additional members within the same union. We also attempted to recruit through organizing meetings, briefly. However, challenges in organized labor due to the current federal administration made \edit{getting interest and engagement} very difficult. \edit{Additionally, we provide a summary of the self-reported demographic data collected from our pre-interview survey in Table~\ref{table:demographics}. Note that participants were able to select multiple identities from our list and that three participants recruited through snowballing did not fill out the pre-interview form.} 

\begin{table}
\small
\setlength\extrarowheight{3pt}
\setlength\tabcolsep{0pt}
\begin{tabular*}{0.95\textwidth}{ @{\extracolsep{\fill}} 
       c c c c @{} }
\hline
\hline
\textbf{Demographic} & \textbf{Group} & \textbf{Percentage} & \textbf{Count} \\
\hline
\multirow{5}{*}{Ethnic Background}
    & Hispanic & 13\% & 2 \\
\cline{2-4}
    & Black or African American & 19\% & 3 \\
\cline{2-4}
    & Native American or Alaska Native & 6\% & 1\\
\cline{2-4}
    & White or Caucasian & 31\% & 5\\
\cline{2-4}
    & Asian or Pacific Islander & 25\% & 4\\
\cline{2-4}
    & Prefer to self-describe & 6\% & 1\\
\hline
\multirow{3}{*}{Gender Identity}
    & Man & 43\% & 6\\
\cline{2-4}
    & Woman & 50\% & 7\\
\cline{2-4}
    & Non-binary & 7\% & 1\\
\hline
\hline
\end{tabular*}
\caption{\edit{Overview of self-reported demographics from 14 of our 17 participants. Participants were able to select multiple identities with percentage being calculated as the number of counted reports over the total number of selected identities.}}
\Description[A breakdown of the collected self-reported demographics]{A full count and percentage estimate of the ethnic background and gender identity of participants}
\label{table:demographics}
\end{table}

\subsection{Interview Procedure}

Our interview questions are available in Appendix~\ref{appendix}. We started with background questions both on the individual and the labor union before moving on to communication. Questions started with lower-level details around structure and usage before moving to higher-level questions around dynamics and culture. We conducted three practice interviews to help inform and validate our initial drafts of the interview guide. These interviews were conducted by the primary author on fellow labor organizers within our university. These interviews helped identify areas where questions overfit to prior literature, leading us to make small wording changes to better encompass relevant interactions.

Interviews were conducted through Zoom and took an average of 53 minutes. During initial email conversations, the primary author provided an informed consent form detailing the study's data collection and usage policy and collected a signed version of the form. At the start of the Zoom call, the interviewer gave a high-level review of the data policy and allowed the interviewee to ask any questions. After this discussion, the interviewer began recording through Zoom to their local machine. Recordings were then transcribed through Rev.com's AI transcription service, \edit{only after we opted the account out of future AI training}. The primary author then read through the transcripts to fix any errors and remove any identifiable information.  

\subsection{Analysis}

We analyzed the data using an iterative open coding approach~\cite{braun_using_2006}. After completing 10 of the interviews, the first two authors reviewed three interviews, highlighting relevant themes and insights within the text. Each author then independently drafted a codebook from their notes before coming together to compare and synthesize the codes into higher-level themes. After discussions, the two authors then presented the codebook draft to the fourth author for refinement. The two authors responsible for codebook development then repeated this process, each reading three interviews and making edits to the codebook. During this step, a few small organizational changes were made, but both authors felt the codebook properly captured relevant themes. Next, the two authors conducted a round of focused coding to test usability. Afterward, the authors combined several sub-codes to help increase usability. Finally, the two authors completed one more round of coding and discussion to ensure a strong codebook. 

Once the codebook was stable, the first author, who is the most experienced with organized labor, independently and systematically applied the codebook to each interview using NVivo 15. During the coding process, the first author met with the second author every 4 to 5 interviews to discuss themes and resolve any issues that arose. After these meetings, the second author audited the coded interviews by thoroughly inspecting any interviews that the primary author found difficult to code and skimming the other coded interviews.

\subsection{Positionality}

As academics and labor organizers, we play multiple roles within the complex societal systems of capitalism, labor, and technology that shape both our access to participants and our interpretation of the data. Many of the authors have significant, multi-year labor organizing experience, with the first author having been involved in multiple activist campaigns and organizing efforts. These backgrounds, along with the position of the academy in labor struggles, likely influenced both participation and our analysis. As fellow labor organizers, we likely had an easier time finding other labor organizers to interview since we recruited through our networks. This shared background also likely made disclosure easier for interviewees, as participants and interviewers already shared rapport and a common perspective, while also easing analysis given our strong existing frame for understanding technology in unionization. At the same time, because the academy often has strained relations with organized labor, workers may have withheld certain sensitive or strategic information from us.

\subsection{Ethical Considerations}

As labor organizing has a well-documented risk of employee and legal retaliation~\cite{mcalevey_no_2016}, we took a number of measures to reduce risks to participants. The procedure was developed following the security protocols used in local labor reporting. First, all recruitment was done outside of workplace channels; that is, all advertisements were posted in union-only channels (to the best of our knowledge). Second, all recordings were done locally to ensure that Zoom had no cloud access to the data. We also ensured that our transcription service, Rev.com, had opted our accounts out of data reuse and machine learning model training. Third, after interviews were completed, the first author went through and removed all identifiable information concerning the individual and the workplace. We did not remove national or parent union information based on our assessment that doing so would not hurt participant or local privacy, and understanding the national union context was potentially important in the analysis. We also stored no links between the participants' information and the de-identified transcript. 

\subsection{Limitations}

There are a few limitations to this study that are important for contextualizing our results. First, participants were only from the United States. Organized labor laws can differ greatly from country to country~\cite{garneau_digitalisation_2023, molina_it_2023}, and thus we feel it would be inaccurate to generalize across nation-state boundaries, especially in the global south. We also want to note that recruitment proved more challenging than originally expected, likely in part due to Trump's attacks on organized labor, including the removal of Gwynne Wilcox from the National Labor Relations Board and the removal of collective bargaining rights for specific federal workers~\cite{mansfield_supreme_2025, oakford_trump_2025}.

Furthermore, by limiting our scope to workers and unions who are \textit{already} using digital communication technologies, we might miss union workers who have experience organizing through digital tools but choose to leave these platforms, \edit{and miss learning about their reasons for opting out}. Additionally, there are limitations that come with conducting interviews in contested spaces regarding what information participants feel comfortable sharing, \edit{as discussed above}. 

\edit{Finally, we wish to mention our lack of focused discussion of race within this work. Recognizing the importance and history of American labor unions requires critical conversations about the deeply troubling and racist history of labor disputes and organizing~\cite{loomis_history_2018}. While race did arise in our interviews---with some participants citing it as a motivating factor for their organizing efforts---we did not directly examine racial dynamics in digital organizing.}

\section{Results}
\label{sec:findings}

In this section we outline how workers and unions structured their digital union spaces, the dynamics that formed around these spaces, and their effects on broader union culture. We also provide discussion on the observed inter-union power dynamics \edit{within the digital spaces studied}. At a high level, we observed workers having a high amount of agency through the use of digital tools, with worker leaders often managing the structure and usage of them. Participants reported these spaces being used for a mix of community building, organizing activities, sharing information, and union democracy, while also running into \edit{critical} challenges related to information overload, increased conflict, and consensus building. Finally, we want to note that all of the unions interviewed 
were heavily worker-led, likely shaping these results. 

\subsection{Constructing Digital Union Spaces}

\begin{table}
\small
\setlength\extrarowheight{3pt}
\setlength\tabcolsep{0pt}
\begin{tabular*}{0.95\textwidth}{ @{\extracolsep{\fill}} 
      lcl @{} }
\hline
\hline
\textbf{Platform} & \textbf{Total Users} & \textbf{Participants Who Reported Using} \\
\hline
Signal & 10 & P1, P2, P3, P4, P8, P11, P13, P14, P15, P17 \\
Discord & 8 & P1, P3, P4, P11, P12, P13, P14, P17 \\
WhatsApp & 8 & P4, P5, P8, P9, P10, P11, P14, P15 \\
Slack & 6 & P3, P6, P7, P8, P9, P12 \\
Gaggle & 3 & P9, P10, P16 \\
GroupMe & 1 & P17 \\
\hline
\hline
\end{tabular*}
\caption{\edit{Total counts and participant breakdown of the digital communication tools used in union organizing.}}
\label{table:platform}
\end{table}

\subsubsection{Workers used many platforms and prioritized accessibility}

In constructing digital union spaces, workers used a wide variety of tools, including Slack, Discord, Signal, and WhatsApp. Signal was the most popular of these tools, with 11 participants mentioning using it in organizing spaces. Discord and WhatsApp were tied for the second most popular, followed by Slack. See Table~\ref{table:platform} for a full breakdown. Participants also often reported using multiple tools during their experience with a union. This could include using multiple platforms at the same time, like with P15, ``\textit{Okay. So we have various different group chats across many different platforms. We've used WhatsApp, we use Signal},'' or P2, ``\textit{We are trying to use Discord, but most of the time people use [Signal]. So there's 200 [Signal] groups, the [Signal] group for everything.}'' In other cases, \edit{unions shifted their primary digital hub over time.} 

\begin{displayquote}
    Yeah, [the union] was Signal and GroupMe. Then after I left, they transitioned to Discord. Before I joined, they were on Discord for a bit and then gave that up and went to Signal and now they're back on Discord. It's just a curse that keeps happening. [P17]

\end{displayquote}

One of the main factors in choosing which platform to use was accessibility, with a majority of workers attempting to let as many people participate as possible. Often, this would manifest in choosing a platform that was already used at work, like in P12's case: ``\textit{we have Slack for union, our specific union, and that kind of works because [employer] itself has a Slack, so it's already a tool everybody who already works at this company would have.}'' Similarly, P7 explained: ``\textit{people are familiar with using it, it's easy to add another Slack instance on your phone or on your computer. So it makes it easier for people to go back and forth.}'' Other times, unions would end up using multiple platforms to ensure all workers could have access. For instance, P8, an active organizer within the union, mentioned using Slack, WhatsApp, and Signal to ensure coverage of all workers. Despite these attempts at inclusion, some participants identified the use of chat platforms as a barrier, with workers needing to download an application to get fully involved with the union.

\begin{displayquote}
     I think the thing with Signal is that it's just hard to use and it's hard to download and if you never download it or you're not online all the time, you're not going to be part of this. I mean, it's meant to support the in-person organizing. I think it's doing a good job with that. But, [\dots] I feel like it's making groups that are maybe hard to access sometimes. [P3]
\end{displayquote}

\subsubsection{Channel structure was quite messy}

Participants also reported these spaces to be incredibly messy, with workers needing to manage the ``\textit{1700 chats we got going on}'' [P12]. Broadly speaking, participants described channels that fell into one of the following broad categories. First, there were working group channels, such as the bargaining team or stewards chat. \edit{U1 also had a number of caucus channels for specifically marginalized identities---including a Black, Indigenous, and people of color (BIPOC) caucus and feminist caucus.} These channels were typically private and required membership in the respective working group for access. Second, there were community-based chats focused on the workplace or shop floor. These channels were generally open to all union members and used for purposes like ``\textit{communicating about grievances, communicating about discipline ... And then also to share some more time sensitive local news broadly or updates about the joint council meeting}'' [P9]. Third, \edit{workers created} planning chats used to coordinate logistics for direct actions, such as ``\textit{getting food to the picket line}'' or ``\textit{scheduling strike shifts}'' [P11]. Finally, there were mutual aid or solidarity channels. These included spaces for asking stewards contract-related questions, watching bargaining sessions, or engaging in non-union discussions like sharing music recommendations.

Participants also stressed the structure of the digital union space as being important to organizing. As discussed, many of these tools tended to become overloaded, with messages coming in too fast for productive discussion. To help maintain functionality, workers would try to separate channels for work and discussion. As P14 said, ``\textit{I think this is why we tried to have different chats for different reasons. So just like if you're going to be in the strike-specific chat, just keep drama out of it}.''

\subsubsection{Security and privacy concerns were both internal and external}

Workers also had a well-developed notion of privacy. Channels discussing information that, if leaked to management, could harm the union tended to be private. The prime example of this was the bargaining team \edit{channel}, as they possessed the most sensitive information with respect to threats from management. This also extended into inter-union communication, with more organizing-focused workers trying to avoid broadcasting sensitive information about co-workers, like potential disciplinary actions or identity issues. 

\begin{displayquote}
    The bargaining committee and stewards channels are private just because a lot of times you'll be handling sensitive information about other people in the union. You need to be able to speak pretty frankly with each other, especially as you're trying to organize individuals. You don't want to offend anyone. And then the outreach channel, since you're organizing big actions, there are always going to be a few people in your union that are anti-union. And so you don't really want them to catch wind of those, but anyone who wants to be involved in those channels can be added to them. We just try and keep it a little bit more shut and sensitive so that it somehow doesn't get leaked out before we try to do a big mobilization or something like that. [P7]
\end{displayquote}

In response to these privacy threats, the unions we interviewed tended to have some sort of verification procedure for getting access to the digital union spaces. In the two more established unions we interviewed, U1 and U3, there was a moderately formal process for getting access to the larger digital union space. In U1, workers who wished to join the union Discord would submit a Google form with their ID number, which was then checked against the list of union members. In U3, workers were introduced to the union-platform during union orientation and allowed to join if they signed a union card. 

\edit{By contrast, workers in newer unions or those using less centralized tools tended to rely on more informal access mechanisms.} For instance in U2, ``\textit{somebody would reach out to somebody and be like, `hey, there's [a] union Slack, are you interested in joining?' And then that person would get added and everyone would be like, `hey, welcome'}'' [P6]. We also saw similar informal and in-person methods used in smaller community-based chats that used a less centralized platform. ``\textit{If I remember correctly, we had a department meeting at the beginning of the strike, and then someone pitched the idea, founding a chat group, and then we said we would use WhatsApp. And then we just passed a sheet of paper and everybody wrote their phone number and names and stuff like that}'' [P4]. 
 
Workers also faced some technical security challenges in creating these digital spaces. For instance, P1 mentioned challenges in maintaining security if an account was compromised: ``\textit{If a cop gets your phone, we have to go through and manually remove you one by one from each of the different texts.}'' The most consistent challenge was accessing digital union channels through work-managed technologies. In about a third of interviews participants mentioned fears over surveillance through company-owned devices. These ranged from simple challenges like using a company email to more serious issues regarding the use of company devices. Perhaps the most concerning example was from P13, who reported co-workers being afraid of the employer hacking their phones or personal emails. While we are unable to verify these claims due to the nature of labor struggles, there was one union, U2, in which workers reported layoffs due to posts within their union's Slack. Interestingly, no workers interviewed reported fears over the platform leaking information, nor their data being used in AI-training, despite the real threats of both~\cite{westerlund_can_2024, team_inside_2024}. This is likely in part due to the obfuscation of technological tools and workplace surveillance, but are also due to the allocation of capital for technological development, as workers often lack the means to access or create digital spaces without privately owned or managed technology. 

\subsection{Dynamics in Communication}

\subsubsection{Digital tools were used for most facets of organizing}

Workers used these digital spaces for a number of different tasks. More public-facing channels would be used for activities like sharing updates, 
asking and answering questions, 
and handling bureaucratic procedures including sharing links, files, surveys, and votes. 
Private channels tended to focus on organizing, with workers planning direct actions,
sharing organizing report-backs 
or doing power mapping. 
This is best illustrated by P9: 

\begin{displayquote}
    We have our local shop on WhatsApp, which is everyone, all [union] members, and we use that for communicating about grievances, communicating about discipline, communicating about sort of real time in the moment \dots
    [The union] is starting a broader WhatsApp group for sectoral bargaining and [\dots] we're eventually going to start a strike prep WhatsApp group, but right now we're calling it CAT, which is the contract action team. [\dots] And then [a] grievances [channel] for the delegates to keep track of grievances and [\dots] more sensitive communication potentially [\dots] about discipline. [P9]
\end{displayquote}

Some participants also described attempts at having organizing conversations digitally. However, doing so proved to be a challenge. Organizing a union requires developing significant trust between workers, especially as there are numerous well-documented instances of large-scale union busting~\cite{logan_corporate_2025, logan_crushing_2021, logan_union_2006}. Developing this trust between workers through digital tools proved a difficult task. P17, who had significant organizing experience, mentioned: ``\textit{A good organizing conversation is one where an organizer listens and can sit in silence and wait for them to respond, but if you can't read them, you are going to feel like you have to play catch up.}'' In U2, where workers were almost exclusively organized through Slack, we heard about this exact dynamic playing out.

\begin{displayquote}
    Because it was in this digital medium that made me feel kind of icky 
    \dots I chatted with them and I tried to keep [\dots] an open mind. 
    \dots I think my reflection is, it is hard to be an organizer in a digital age in particular because there's less ability to form human connection upon which to build strategy. You kind of just have to go straight for strategy. And that was my experience being a recipient. [P6]
\end{displayquote}

In response to these challenges, organizers like P13 started using more creative methods of reaching out. 

\begin{displayquote}
    You would be like, `hey, we've never talked, do you want to do a coffee chat?' Or, `hey, I saw you posting in the [game] channel that you're a big Deep Rock Galactic person. Do you want to meet up and play?' Just trying to find different ways to get people off work platforms and having conversations about their working conditions. [P13]
\end{displayquote}

Finally, in about half of the interviews, we observed these digital spaces serving as spaces for union democracy.
The most common form of this was for smaller organizational votes. For example, in P11's interview, they mentioned using WhatsApp's polling feature for gauging worker sentiment over small decisions and using emoji reactions for RSVPing to a strike action. These use-cases tended to be well-received, with workers being able to quickly take a temperature check and plan better. However, when engaging in more controversial, or as P14 put it, ``serious things,'' workers started to run into issues.

\begin{table}
  \small
  \centering 
  \begin{tabular}{>{\raggedright\arraybackslash}m{0.3\linewidth}|>{\raggedright\arraybackslash}m{0.3\linewidth}|>{\raggedright\arraybackslash}m{0.3\linewidth}}
    \hline
    \hline
    \textbf{Challenges in Organizing} & \textbf{Technical Component} & \textbf{Broader Effect} \\
    \hline
    Issues of accessibility and ensuring that all workers who are interested have access to digital spaces & Worries over retaliation through workplace surveillance and the need to use a non-familiar application & More organizing-focused workers having to manage multiple platforms, increasing workload. \\
    \hline
    Worker-on-worker conflict was reported to be amplified, especially when handling challenging or controversial decisions & Digital, text-based tools reduce the ability for fellow workers to see the impacts of their words or actions & Workers would try to avoid using digital tools for discussions, especially controversial ones, and develop forms of verification for access to the digital union space. \\
    \hline
    Worries over privacy, both collective---with respect to the union---and individual with respect to interpersonal issues between workers & Technical tools could increase the visibility of sensitive information, leading to management or other workers taking action or issue & More active and organizing-focused workers would use private channels to avoid information leakage. \\
    \hline
    Challenges in building a large consensus or deciding what actions to take & Technical tools allow for workers to share opinions without having to engage with other workers & Fears over workers not understanding what it means to unionize along with shifts in participation levels. \\
    \hline
    \hline
  \end{tabular}
  \caption{Summary of identified technologically driven challenges in organizing digitally}
  \Description[A three column summary of our results]{A three column summary of our identified challenges in organizing, the technical component, and its broader effect}
  \label{table:issues}
\end{table}

\subsubsection{Challenges included conflict, overload, and moderation}
\label{sec:findings:challenges}

Participants mentioned a number of significant challenges in organizing digitally. Table~\ref{table:issues} shows a summary. Part of these challenges stemmed from the tendency for digital union spaces to become centers of worker-on-worker conflict, 
with participants describing their Discord as ``\textit{intense}'' [P2], ``\textit{toxic}'' [P11], and causing ``\textit{bad blood}'' [P15]. For instance, in U4, discussions over striking led to a public fight between two workers, in which one of the non-union-supporting workers was called out publicly for flashing their wealth on Instagram. This led to a number of internal fights and arguments between workers on the Discord, which, according to P11, ``\textit{still hasn't really died down.}'' As the fighting continued, all of the participants we spoke to ended up leaving the Discord, including one of U4's Discord ambassadors, P14, citing the constant infighting and potential security leakage.
\edit{This trend of conflict between workers being amplified in digital spaces was consistent across all the other unions in which we interviewed multiple workers.} 
For example, P6 mentioned a public email chain in which, ``\textit{A coworker of mine who was skeptical of the union, who I was interpersonally close with, replied to an all staff email with some critique of the union.}'' One of the public union leaders saw this email \edit{and responded} to it, which ``\textit{turned up the heat.}'' This lead to a flurry of union Slack activity, both public and private, and a significant amount of stress upon P6, who reported ``\textit{not sleep[ing] well that night.}''

Another main issue was information overload. A majority of workers found these digital spaces to be too noisy, crowded, or overwhelming. 
As P14 put it, ``\textit{If I don't pick up my phone for an hour, I'll miss a hundred messages.}'' This problem was especially prevalent among workers who used Signal and WhatsApp.
Without the ability to pin messages or create threads, workers found it easy to lose the important details of the conversation. Small planning conversations with a few people could quickly become 30 or 40 messages, making it easy for other organizers to lose track of the conversation. Workers would work around this by using features like message reply. As P15 summarized, ``\textit{I can hit reply and it's going to bring the message back to the bottom with my reply to it and the conversation's then able to continue.}'' Other times workers, like P8, would start using multiple platforms, one for discussions with workers and the other for sharing and organizing files.

This lack of message organization would then compound, with workers feeling the need to read every message. Perhaps the worst of this was with P2, who reported: 

\begin{displayquote}
     I think I've gotten better at, like, scrolling through through my Signal and being able to, like, mentally filter it out and, like, make sense of it. But it's, it is overwhelming. I remember [\dots] when I became, like, really involved with the union organizing, that was one of the most like, overwhelming thing[s \dots] Just how many chats I was part of, and how hard it was to keep track of everything, how compulsively I [\dots] I need[ed] to check, like I've gotten better about it, but yeah. [P2]
\end{displayquote}

This tendency for worker-on-worker conflict and information overload would then shift workers usage of channels. \edit{Workers who were less active in the union---such as P6 and P10---chose to disengage in response.} For example P6 ``\textit{did not really engage}'' with the union's Slack, as they were ``\textit{afraid of being called out.}'' \edit{Conversely}, many active workers would instead move to smaller or closed channels of communication. 
P9, a worker whose union had significant infighting on their public channels, mentioned shifting to the much smaller and more local WhatsApp chats, and P8---a worker in the same local---said they reduced the number of platforms to just one: ``\textit{So what I was saying is, with our three platforms that we were using, which was Slack, WhatsApp, and Signal, there came a point in time where we decided that we needed to narrow it down to just one}'' [P8]. 

Workers also tried to help alleviate some of these issues through light forms of social moderation.
In unions U1 and U2, there had been some discussion about creating formalized rules; however, such discussions were rare. P1, who was responsible for moderating their union's Discord, mentioned having some very light moderation regarding hate speech, and P3 said the rules were ``\textit{don't be an asshole or something.}'' P6 also mentioned having some discussions about Slack usage, but that they weren't enforced. One common strategy was to have some of the union leaders pull more heated individuals into private chats or calls and discuss it one-on-one. 

\begin{displayquote}
    I think most of it was like, `hey bud, it seems like you're a little bit heated. Can we talk about this? Can we just do a quick call? Can we just hop on a call? Let's chat through this.' And I think that was difficult because a lot of people took that as like, well, why are you saying I can't talk in public? And it's like, no, no, nobody's saying that this is going to be a more productive conversation because I can listen to you here. There's not 15 people chiming in. It seems like we could just chat this through and I think that is the correct thing to do. [P13]
\end{displayquote}

\subsubsection{Consensus building was extremely difficult}

Perhaps the most significant challenge of organizing digitally was consensus-building. In almost every interview, participants mentioned challenges in building consensus.
As P3 puts it, ``\textit{I think when trying to make decisions on these platforms [it] is not as fruitful just because, I dunno, you're less able to do the consensus building, which I think is really important for running a participatory, I dunno what to call it, organization thingy like a union tries to be.}'' Messaging platforms like Discord and Signal allow workers to share their opinions on organizing efforts instantly. When compared to in-person discussions, in which only one person can be talking, workers are able to share their opinions without having to listen to or meaningfully engage with other arguments. This potentially poses a challenge to digitally-facilitated organizing. P2 reflected on this in their interview:

\begin{displayquote}
    For example, a lot of grad workers who are maybe new to the idea of a union, think of a union as primarily a service that they're signing up for, versus an endeavor that they're like, a project that they are joining. They will like, I don't know, argue for more asynchronous decision making... we had this whole conversation about why that would actually be really bad for organizing, because then people would develop this idea that you can just make a decision by clicking on a link. That's not what decision-making is. Decision-making is developing consensus and collective will. That is why we should make decisions together in a meeting where you can actually talk about it and like, come to a shared consensus like that is the goal of decision making. [P2]
\end{displayquote}

All of this led to many of the workers interviewed 
to conclude that, while these tools were helpful, they were not a substitute for real in-person organizing. Both P1 and P13, who were the most positive participants regarding digital organizing, mentioned the importance of one-on-ones: ``\textit{I think like for more of an organizer perspective, \dots use all the tools at your disposal, but also like one-on-one conversations tend to work the best}'' [P1], and ``\textit{These are tools, I think to enhance your organizing efforts, augment it, but they don't replace the building blocks of conversations, of one-on-ones, of outreach, of all of that stuff}'' [P13]. 

\subsection{Cultural Impacts of Digital Technology}
 
\subsubsection{Visibility and humor increased worker solidarity}

One significant impact on union culture was worker solidarity. Digital tools have a well-documented history of shifting who or what is visible~\cite{asad_illegitimate_2015}. A prime example of this is in U4, \edit{where workers were split between customer-facing and kitchen staff, with little contact between the two groups.} Access to these digital spaces allowed workers, who had never met face-to-face, to help build community with each other. P14, who worked in the bottom bar, a back-of-the-house and often isolated station, mentioned not ``know[ing] most of the people in the chat.'' However through these digital union hubs, he was able to build solidarity, commenting, ``\textit{So I'm tight with them in the chats kind of, if that makes any sense, even though I never talked to them in real life.}'' A significant portion of this community-building was through shared humor or memes.
 
\begin{displayquote}
     We're just silly [\dots]
     there's a difference between being [an] automated computer fighting for the cause, and we are a community [\dots] 
     We still get to have a little fun and goof around in our little chats and stuff, and that eases tension that lets us know that we're just people that live in our lives and doing things, and that just happened to be trying to take down a horrible, horrible practice of abusing workers. That's what this is worth. That's what all this is for, to be able to just send memes and communicate with each other and have a laugh. [P15]
\end{displayquote}

Another great example of increasing visibility to help build solidarity was through the use of ``bargaining observer chats.'' While not always available---often due to management's refusal---workers can sometimes get management to agree to open bargaining. \edit{Unlike closed bargaining, where management and the union's selected representatives meet behind closed doors, opening bargaining allows for all members of the bargaining unit to attend and observe}. Workers in U1 and U2 set up a channel in their Discord and Slack, respectively, for workers to instantly share frustrations over their public bargaining sessions or changes in the workplace. ``\textit{And people could live chat in that channel through bargaining and through all-hands meetings. So we had places to talk through shared events, and those were really, really lively because you would get people being like, what the fuck? Have these people ever paid a mortgage?}'' [P13]. 

\edit{This seems to suggest that chat-based tools are a good way to help facilitate some degree of community-building by increasing the visibility of worker struggles and the sharing of humor.} P17, who is in the process of organizing workers across a city, shared the best example of this: 

\begin{displayquote}
    Unless we did that deliberate work of bringing people together, it's very easy for each hall to stay its own hall and not talk to each other. We wanted to combine those and bring those together because sometimes one hall's conditions are significantly worse than another. Sometimes their treatment or their living conditions are significantly different, and you want to build up that sense of solidarity by reminding that everyone's in this community together and you want to bring the floor up for everybody together. [It] helps with people wanting to be more invested in improving conditions and also getting to know each other. [P17]
\end{displayquote}

\subsubsection{Organizers had fears over shallow organizing}

While these digital union spaces were helpful for building a community, we observed a number of fears that these digital spaces would promote a shallower union\edit{, one lacking the strong relationships needed for successful workplace organizing}. Forming and maintaining a strong union requires significant amounts of labor; there are ``no shortcuts'' to deep organizing~\cite{mcalevey_no_2016}. In order to be successful, workers need to take ownership within their union, not see the union as an ``other'' or third-party, but as a representation of their own voices. Workers also need to develop a strong sense of solidarity and take action as a collective. 

These fears were also compounded by the increased ease of interacting with union democracy, leading workers to be afraid of promoting ``service-unionism'' through their \edit{use} of digital tools. P9, a shop-steward in U3, reflected on this, saying, ``\textit{there are really great applications, but it can also make it easy to stay at home and [send] an email and then not show up to places, I think.}''  For example,  P16, our least active participant, reported reading messages but not participating despite supporting the union, citing a lack of personalization and the ease of ignoring them.

In a few instances, workers found clever workarounds, utilizing the tools as a place of contestation to help organize. P12 shared the best example of this, with workers using the union Slack for announcing grievances\textemdash contract violations usually done by management. By using the Slack to highlight how management was breaking their current contract, organizers were able to triple the number of people participating in organizing discussions. Workers then created a thread highlighting the grievance on the workplace Slack, with many of the people who helped organize the action providing reaction emojis to show their support. Other workers who had not been as active in the union saw the thread take off and also gave supportive emojis, with the original message reaching a Slack-imposed reaction cap. This seems to suggest that platforms like Discord and Slack \textit{can} be used in militant organizing; however, doing so requires an adaptation from the more standard union play-book~\cite{blanc_worker--worker_2024}. 

\subsubsection{Digital spaces became a source of authority}

Finally, we turn to leadership. In interviews, participants reported these digital union channels becoming a source of authority themselves. Because these platforms became the hub of organizing, \edit{the information and interactions that happened on them became more ``real'' than those elsewhere}. Organizers found that certain actions, those that were critical for the union to function, needed to happen through digital channels in order to have authority. Organizers also reported using these tools as a way of \textit{establishing} authority. For instance in U1, participants mentioned controversies over the union being anti-democratic and failing to engage with the full worker base. In one instance, union leaders claimed they had engaged with the broader worker base by posting in union channels, implying that union leadership equaled digital channel usage. Another example is P13, who felt that shifting the platform from a less organized tool, Signal, to a more centralized and regulated one, Discord, was important for demonstrating that their union drive was ``official.'' P17 also noticed a similar phenomenon when organizing, noting that ``\textit{oftentimes when you're talking to people that are not a hundred percent convinced, you're always going to need to direct them to a source that is of a higher authority than your own. So it's like, here's the link to the Instagram post, [this] is real, [it's] happening.}'' In summary, digital channels used for organizing became a source of leadership unto themselves, with workers seeing their usage as being tied to their sense of what a union is.

\subsection{Digital Power and Control in Union Organizing}

\begin{table}
\small
\centering
\begin{tabular}{>{\centering\arraybackslash}m{0.3\textwidth}>{\raggedright\arraybackslash}m{0.65\textwidth}}
\hline
\hline
\textbf{Type of Chat Tool} & \textbf{Effect on Power Dynamics and Tensions} \\
\hline
\makecell{Centralized tool: \\Discord or Slack} &
\begin{itemize}[leftmargin=*, noitemsep]
  \item Workers served as administrators by directly controlling platform structure.
  \item This sometimes led to worker-to-worker conflict over platform administration.
  \item Workers had easy access to channels outside of staff purview, with explicit structure to facilitate them.
\end{itemize} \\ 
\hline
\makecell{Decentralized tool: \\WhatsApp or Signal} &
\begin{itemize}[leftmargin=*, noitemsep]
  \item Workers and staffers shared administrative access to digital channels.
  \item No workers reported conflicts over administration or access between workers and staffers.
  \item Workers had access to channels outside of staff, though they reported not needing them.
\end{itemize} \\
\hline
\hline
\end{tabular}
\caption{Summary on the choice of tool used for union communications and its effect on the power dynamics within the union.}
\Description[A table summarizing the results of this section]{A table summarizing the choice of tool and its effect on the pwoer dynamics within our studied unions}
\label{table:powerDynamics}
\end{table}

In our interviews we observed two main types of power dynamics within the unions' digital communication hubs. When workers used a more centralized tool like Discord or Slack, workers were directly in control of the platform.
When workers were using a less centralized platform like Signal, workers often split management between workers and staffers.
Table~\ref{table:powerDynamics} summarizes these results. In unions 1 and 2, workers were in direct control of the digital spaces, allowing them to maintain ownership over organizing. \edit{While in unions 3 and 4 workers often lacked formal administrative control, the tools they used provided them with easy ways to exert control, allowing them to create channels that centered workers. Critically, and in all cases studied, we found that the use of digital text-technologies helped increase worker agency and shifted organizing to be more worker-focused.} In more orthodox methods of organizing, workers \edit{typically} rely on full-time staffers to bridge spatial and temporal gaps, \edit{serving as} the central communication point for union affairs~\cite{blanc_we_2025, pasquier_democratic_2019}. However, as our earlier sections showed, many of the critical aspects of labor organizing were shared or handled through these digital platforms. This represents a shift in labor organizing practices, helping to keep workers at the center of labor organizing.

\subsubsection{Workers had direct control over communication} In U1, workers had established a committee with elected officers to manage the Discord. In our interview with one of the co-chairs, she made it clear that the tech committee was responsible for handling all facets of the union Discord, including adding members, managing channels, and responding to security threats. ``\textit{Yeah, so [staff] don't have a ton in terms of like responsibilities with digital communication usage other than like they participate in it, really. It's mostly tech's job}'' [P1]. She also made no mention of staff attempting to manage the platform or workers' usage of the platform. There were, however, some conflicts between workers: ``\textit{previous officer cores wanted
to get all the admin privileges and then they wanted to like implement a bunch of changes to the Discord
and basically the previous tech chair said like, no, get out of our business}'' [P1]. However, in both her and the other interview evaluations, these were not destructive conflicts, with no workers disconnecting as a result. P1 and P2 did, however, bring up numerous issues with the parent-union-managed database, leading them to create their own managed database, which reportedly ``\textit{helped organizing a lot.}''
  
Union 2 had a significantly less developed infrastructure for control of the platform. Critically, however, workers were the ones in charge of their Slack and occasionally used it to create channels without staff involvement. Access to these channels allowed workers to \edit{determine the best ways} to respond to issues relating to union staff.  

\begin{displayquote}
    For the contract bargaining committee, we had a separate CBC channel that was just the five of us. And so it was really important [\dots]
    that we had that space to be like, what the fuck just happened? Why did they make that decision without talking to us? Or what gets a vibe check with each other without our lawyer and our organizers seeing that information, you need to be able to have frank conversations with each other too. [P7]
\end{displayquote}

One of the staffers we interviewed, P17, also shared a similar reflection highlighting that there are ``\textit{shit staffers}'' out there and that workers ``\textit{need to find ways to circumvent them}.'' 

\subsubsection{Control over communication was distributed}
In comparison to unions U1 and U2, unions U3 and U4 had little to say about issues relating to their parent union. Part of this, likely, is driven by \edit{their membership} in the United Auto Workers Union (UAW), which has recently undergone significant bottom-up reform. Both unions' primary tool for communication was WhatsApp, which allows for easy channel creation and member addition. In U4, channels were usually created by the bargaining team or active organizers within the union. Workers would also often add their main staffer who helped with organizing, but workers reported \edit{that they were not in all of the chats}. In U3, staff access to chats tended to vary depending on the channel, with workers and staff administration being split between the two groups. In one instance, P11 was surprised to find that the staffer was responsible for managing their workplace channel. 

Instead, participants \edit{reported that staffers were mostly there to answer questions about union by-laws} or to provide a more knowledgeable opinion. One of our participants even described having access to direct communications with the national union. P8 mentioned having a digital communication channel that allowed ``\textit{leaders from the president}'' all the way down to ``\textit{rank-and-file members}'' to communicate with the parent and national union. Access to these channels aided P8 in properly navigating a potential disciplinary action. More broadly, participants mentioned very little about needing to worry about access or internal power structures, likely shaped by the ease of creating worker-only channels within the tools being used.

\section{Discussion}
\label{sec:discussion}

Our findings reveal a number of larger design, structural, and cultural challenges present in digital labor organizing \edit{that have significant impacts on how we understand and theorize about the role of technology in union organizing. Our conversation focuses in on two threads: the (im)possibilities of digital organizing and the role of digital tools in union renewal. Importantly, we do not give design recommendations. Many of the digital tools were selected for their ease of access and are developed by privately owned companies unlikely to care about a labor organizers' needs or wants, and thus articulating design recommendations seems unproductive. Instead, we give reflections on the potential role of these digital tools in aiding and augmenting organizing efforts in our conversation on union renewal.} 

\subsection{Understanding the (Im)possibilities in Digitally-Mediated Organizing}

\edit{While digital tools are often framed as novel solutions to longstanding problems of coordination, communication, and scale, our results suggest a more complicated picture. Rather than resolving core challenges of union organizing, digital communication technologies frequently re-articulated familiar tensions around trust, authority, inclusion, and worker power. In this section, we place our findings in conversation with prior work on social computing and organizing, strengthening and extending prior theories on the role of digital communication tools in organizing broadly. This conversation also highlights that the effects of digital communication are neither uniformly enabling nor constraining, but instead reflect longstanding struggles over power, legitimacy, and collective action in new technological forms.}

\subsubsection{\edit{Successful digital organizing needs prior common ground.}}
\edit{In our interviews, workers highlighted both successes and failures in using digital tools for organizing and solidarity-building. For example, workers like P6 reflected on feeling ``icky'' when having online organizing conversations, while workers like P13 and P15 highlighted community-building through the sharing of humor and frustration. A natural question, then, is what contributes to these varying outcomes? Building on prior work and the context of our results, we argue that digital organizing is more likely to succeed when common ground is already established. In their classic paper ``Distance Matters,'' authors Olson and Olson argue that, despite significant strides in digital technologies, remote work and collaboration will not be able to replace in-person collaboration. Drawing on organizational and communication theory, the authors identify four integrated concepts that serve as barriers to replacing in-person work arrangements with remote and digital ones---with the first of these being the establishment of common ground~\cite{olson_distance_2000}.} 

\edit{As they define it, \textit{common ground} refers to the shared understanding of what is common between two people communicating. It is in part constructed by the cues given throughout conversation and is essential to more fully communicate~\cite{olson_distance_2000}. This plays out in two main ways in our results. First, we theorize that access and participation within internal union communications help to establish common ground. Examples of this include the workers in U4 who shared humor in their chats and the workers in U1 and U2 who utilized bargaining observer chats to share frustrations. In these instances, access and participation communicate to other workers that one is engaged in and supports worker organizing, allowing workers to communicate more freely and build community. Second, we see it in the attempts to establish common ground in early organizing conversations. As our results highlight, workers attempting to organize new members digitally would often try to find a non-work-related shared interest before having an organizing conversation. Organizers also reported difficulties due to a lack of social cues in conversation, like P17 who highlighted the inability to observe or read reactions, another component of common ground~\cite{olson_distance_2000}. This seems to suggest that (a) understandings of social computing in work contexts overlap with the social computing problems in labor organizing---especially around communication practices---and (b) that despite over twenty-five years of research, previously identified problems around social computing and communication are still prevalent across previously unexplored or understudied domains~\cite{bjorn_does_2014}.}   

\subsubsection{\edit{Platform selection for familiarity can promote workplace technologies}}

Consistent in our results was the focus on accessibility and inclusivity, especially by union organizers when deciding which platforms to use and how to use them. Prior work on digital tools for movement-building by Ghoshal and Bruckman~\cite{ghoshal_design_2019} found that, while organizers had a strong focus on inclusion, the choice to embed digital tools like Slack in organizing could alienate or disempower members, as they were required to become familiar with Slack for full participation. Ghoshal and Bruckman then call for organizers to be aware of the potential barriers that computational technologies pose in organizing spaces.

In contrast, we found that organizers in our study \textit{were} aware of these barriers and often attempted to overcome them by choosing tools used in their respective workplaces. While this promotes inclusivity because platforms are familiar to workers, using platforms designed for workplaces creates new tensions. In particular, many of these platforms are designed for employers, with established norms around internal organization and usage. As these patterns are established through workers' experience with them in the workplace, workers may then seek to replicate similar dynamics or understandings of how to use these technologies in organizing. We can see this in P2's reflection on how new graduate workers can see the union as a service, something that one signs up for, and not a project with different expectations around participation and activity. Other examples include the fears around shallow organizing, in which organizers worried about these tools creating a union lacking strong relationships or implying that sending a message was active participation.

\edit{We see this as an important result for extending our understanding of workplace-union technological practices. American employers are often understood to have almost total control over the workplace, with unions traditionally seen as a counter-balance~\cite{logan_consultants_2002, logan_union_2006}; however, our results suggest that employers' control over the labor process also influences labor organizing. This may make adopting new platforms and establishing new patterns of use more difficult, even if the platform structure does not fully serve the unions' goals. Thus, we see that even when workers act with clear commitments to inclusion, they must operate within technological terrains already shaped by employers.}

\subsubsection{\edit{Platforms remove friction and pathos that is critical for consensus-building}}

\edit{Our results also reveal a core tension between the communication logics embedded within the digital communication tools used and the organizing practices of workers. Workers in our study used these digital tools for many common organizing practices with varying rates of success. In addition to more direct communication practices, workers also used these platforms for sending files, answering questions, and sharing report-backs or time-sensitive information. For these uses, our participants shared few problems, suggesting that the tools were effective for more bureaucratic uses and information sharing. Workers were also quite positive about using digital tools for small, low stakes votes. Drawing from literature on workplace communication and online communities, these results make sense as many of these bureaucratic tasks are in line with the expected uses of the tools in workplaces settings~\cite{wuersch_digital_2023}.} 

\edit{However, practices less common in workplaces revealed further tensions. The prime example of this was consensus-building, in which almost all participants highlighted difficulties. Standard digital communication philosophies focus on streamlining communication, allowing multiple people to communicate at the same time and across spatial or temporal gaps. However, these design goals contradict the necessary friction and pathos in consensus building and union organizing. Workers no longer need to listen to fellow workers during a discussion nor are there practical limits on the number of people who are able to speak at once. Additionally, our participants described how digital forms of voting can aid in the understanding of a union as a service rather than as a shared project, reducing worker participation in union actions. In this way, the blending of organizing practices with digital tools may actually conflict with the formation of collective will and power.}  

\edit{Similar examples have been observed in other organizing work. Ghoshal et al.~\cite{ghoshal_design_2019} discussed the lack of personal pathos in emoji-based voting and Nguyen et al.~\cite{nguyen_it_2025} found that the adoption of technology to streamline mutual aid work could undermine the relationship-building and trust essential to organizing. In both our results and these papers, integrating digital technologies into critical components of the organizing work can conflict with these organizations' goals and values, while the use of platforms for the bureaucratic components of organizing---sharing files, agendas, answering questions---did not give rise to significant controversy. Recalling McAlevey's definition of organizing as relying on relationship-building to build power, it is exactly these actions that serve as the foundation for organizing. This suggests that attempts to embed or develop technological tools that augment the relational processes that are essential to organizing may produce more problems than they solve.}

\edit{This is \textit{not} to say that there is no value in attempting to create or implement mechanisms and technologies for helping democratic procedures within organizing. Indeed, there are many examples within the history of American unions in which class, ethnicity, and gender have been used to exclude those with marginalized identities. The implementation of features such as private voting or discussion mechanisms may help marginalized people feel safer in sharing or speaking out~\cite{loomis_history_2018}. We see this to some extent in our own results, with some marginalized identity groups creating their own private channels to ensure a safe line of communication. Thus, our call is for practitioners to be sensitive to the dynamics at play, as many potential technical decisions can carry obfuscated sociopolitical ramifications that conflict with union organizers' goals.}

\subsection{Union Organizing and Digital Communication Tools}

\edit{We now turn to the role of digital communication tools within the more formalized, regulated, and bureaucratized organizations of American labor unions.}

\subsubsection{\edit{Recognizing the worker-centered power of digital communication}}
\label{workercenteredtech}

\edit{When answering our second research question, we observed two main types of power dynamics. When workers used a centralized tool like Discord or Slack, they maintained full platform administration, creating channels without staff involvement. In contrast, when using a decentralized tool, workers and staff tended to share administrative access. We did not hear these workers expressing a strong desire to communicate outside of staff awareness, though they noted that they had the ability to do so. While there are almost certainly more factors than just the usage of digital communication tools contributing to these results, we theorize that the use of these platforms helped to center workers within their union organizing.} 

\edit{American labor unions often employ professional organizers who help to maintain and organize for the union, and while these staffers often have good intentions, the heavy use of staffer organizing can often lead to workers becoming sidelined in American unions~\cite{mcalevey_no_2016, bradbury_how_2014}. Labor scholars Michels, Greene, and Grieco argue that the adoption of digital technologies may help solve some of the problems of union democracy. They identify four distinct ``forces'' that negatively impact union democracy: (1) inequality of knowledge, (2) differential control over the means of communication, (3) time, energy, and space poverty, and (4) an uneven distribution of communicative skill~\cite{greene_commentary_2003}. Our results provide strong evidence that the adoption of these digital channels did counteract these forces to some extent. For instance, workers had direct control over the communication tools being used to organize, and in U1 and U2, needed to leverage that control to communicate without union staff. We also observed this shift contributing to the neutralization of knowledge imbalances, as much of the critical information, including bargaining updates and organizing report-backs, was shared through these digital hubs. In fact, workers reported suffering from an information overload, suggesting that while there may be challenges in parsing information, they are likely not suffering from a lack of knowledge regarding union organizing. Organizers were also the ones maintaining these channels of communication and used them to send updates, allowing them to build communicative skill and counteract the previously identified forces.}
  
\edit{Our findings also extend recent labor and design scholarship. Returning to Eric Blanc's work on the role of digital tools in labor activist communities, Blanc has argued that the use of digital tools has augmented traditional union models and promoted a ``worker-to-worker'' model driven by online worker communities. While our analysis presents a more cautious optimism, we find one of his core arguments---that workers are better centered in digital organizing---to be consistent with our findings~\cite{blanc_we_2025, blanc_worker--worker_2024, blanc_emergency_2024}. We also note that this is similar to the recent analysis done by Thuppilikkat et al.~\cite{thuppilikkat_union_2024} on the role of digital tools in organizing the Kolkata App-Cab industry.}

\subsubsection{\edit{Digital union renewal}}

\edit{In the background section, we discussed the concept of ``digital union renewal,'' a loose formulation of ways in which academics, policymakers, and labor organizers could reinvigorate and strengthen labor movements. Returning to these failed attempts at digital union renewal~\cite{waddington_e-communications_2014}, recent scholarship on the topic~\cite{hennebert_what_2021}, and the results of our work, we can now develop a better understanding of the role technology can play in union renewal. As Henneber et al.~\cite{hennebert_what_2021} point out, technology is not a monolith. Workers within our study highlighted both successes and failures, and our analysis revealed that many of these outcomes were shaped in subtle ways. Regarding union renewal, we should not overstate the potential of, nor promote reliance on, digital tools. While workers did report that the technical tools were helpful, the consensus was that organizing digitally was \textbf{not} a substitute for in-person organizing. The overly optimistic claims that embracing digital communication tools would \textit{solve} many of the problems facing unions were not grounded in either an understanding of union identities or the critical relationship-building work needed to sustain labor organizing.}

\edit{Instead, we want to articulate a worker-forward approach to labor organizing and digital technologies. As sociologists Micah Uetricht and Barry Eidlin point out, ``What is needed for labor's revitalization is not an immediate shift in the legal and economic terrain,'' but instead, ``the focus needs to be on building a workers' movement that has the power to create crises and use disruption as a source of dynamism''~\cite{uetricht_us_2019}. While numerous works exist on developing tools to correct the information imbalances within labor struggles~\cite{calacci_bargaining_2022}, providing workers with a space for community~\cite{irani_stories_2016, irani_turkopticon_2013}, or understanding the data practices used in organizing~\cite{pei_for_2024,khovanskaya_bottom-up_2020}, little work exists on the disruption aspect of organizing. Workers possess structural power; yet, we see little conversation about how best to amplify or active it.}

\edit{Building on this gap, we argue that debates on union renewal are not simply about informing or connecting workers, but about recognizing the role of digital technologies in the collective processes through which worker power is built. Our findings suggest that technologies play an important role within future union organizing but can also conflict with the goals of organizing and the development of worker power. Taken together, this points toward a worker-centered approach to digital union renewal, one that treats technology not as a solution to organizational decline but as a contingent and political intervention whose value depends on how it is embedded within ongoing practices of relationship-building, consensus formation, and collective disruption. Rather than asking what technology can do for unions, our results suggest that the more generative question is how workers can shape technology to serve the kinds of movements they are trying to build.}

\section{Conclusion}

Following the Starbucks' workers attempts to unionize their Buffalo store, Starbucks managers sent a letter to all employees urging them to vote no, citing the union as a ``third party'' that would stand between management and the employees. The reality couldn't be further from the truth: workers at Starbucks did everything in their power to keep their organization worker-focused, both through numerous one on one conversations and through their use of digital technologies~\cite{blanc_we_2025, canella_networked_2023}. What our research shows is that this picture is far from simple. Workers need to navigate a number of problems that often go beyond simple communication. As designers of technology, it is our job to support workers in their fight for economic liberation, but to do so we need to understand the workers who are taking charge of their liberation, helping them to build power and then use it.

\begin{acks}
The authors thank: BUGWU for inspiring the project and fighting for a better university, along with Audrey Ballarin, Lucía Vilallonga, and Jacksyn Bakeberg; the political economy and computation reading group, who are all fantastic scholars and have helped this work immensely. Jeff Uehlinger from 509 for helping us find contacts and providing valuable feedback; Andrew Elmore and Ngozi Okidegbe for giving fantastic scholarly insight; and every single organizer we interviewed who continues to fight the good fight. Solidarity forever!
\end{acks}

\bibliographystyle{ACM-Reference-Format}
\bibliography{references}

\newpage
\appendix

\section{Interview Questions}
\label{appendix}
\subsection{Introduction and Background Questions}
\begin{itemize}
\item Can you tell us about where you work?
\item What can you tell us about your union/labor organization?
  \begin{itemize}
  \item
    What groups does your union represent at your workplace and how
    large is it?
  \item
    How old is your specific bargaining unit?
  \item 
    What is the current status of your bargaining unit?
    \begin{itemize}
    \item  
      Currently under contract, or in bargaining?     
    \end{itemize}
  \end{itemize}
\item
  How are you involved within the union/organization?
\end{itemize}

\subsection{Questions on Union
Structure}\label{questions-on-union-structure}
\begin{itemize}
\item
  What are some of the reasons you got involved with the union? Are
  there any specific issues that you are super passionate about?
\item
  When it comes to your union, what are the issues your union tends to
  care about most?
\end{itemize}
\subsubsection{Question for non-leaders}
\begin{itemize}
\item
  Can you tell me about your last contract bargaining round
  \begin{itemize}
  \item
    How involved/connected were you when it comes to bargaining?
  \item
    How do you feel about your last round of contract negotiations?
  \end{itemize}
\item
  What do you know about the people responsible for bargaining?
\item
  Can you tell me a bit about how your bargaining/leadership team is
  structured?
\item
  When it comes to the process for ratifying the contract, what do you
  know?
\end{itemize}
\subsubsection{Questions for union leaders}
\begin{itemize}
\item
  Overall, what can you tell me about your last round of contract
  bargaining?
  \begin{itemize}
  \item
    Do you have opening bargaining sessions?
  \item
    How do you feel about your last round of contract negotiations?
  \end{itemize}
\item
  What do you see as your job is, as a member of your
  union\textquotesingle s bargaining team?
\item
  How do you and the rest of the BT go about determining what to try and
  push for at the table?
\item
  Can you tell me a bit about how your bargaining team is structured and
  the process for becoming a member?
\item
  What are the systems that members who do not possess formal power can
  use within your union to enact change?
\item
  Do you have professional staff members in your union and if so what
  role do they fill within your union?
  \begin{itemize}
  \item
    Have you had many interactions with staff and in what capacity?
    \end{itemize}
\item
  When it comes to the ideology of the union, does your
  union have any official or unofficial backing ideology, and have there
  been any discussions around said ideology?
\item
  Are there any other important areas in terms of your
  union\textquotesingle s structure that you think are important that
  haven't been covered, and if so can you elaborate on them?
\end{itemize}

\subsection{Question of Digital Technology \&
Structure:}\label{question-of-digital-technology-structure}
\begin{itemize}
\item
  What types of digital communication technologies are you using for
  communication and discussion within your union?
  \begin{itemize}
  \item
    How many people are in them?
  \item
    Are you aware of different digital communication technologies being
    used within different sub-organizations/groups within the larger
    union, if so what are these technologies?
  \end{itemize}
\item
  What are the primary tasks the unit uses the platform for?
\item
  How are the communication channels structured/organized? 
\item
  When it comes to {[}platform{]}, are there any features you or your
  union have found particularly helpful?
\item
  When it comes to {[}platform{]}, are there any features you or your
  union have found particularly hurtful?
\item
  When it comes to who controls the means of communication, like the
  ability to add and remove people, send messages, or ping who has
  access to these powers?
\item
  (Yes - Staff) Does staff have access to these channels? Why or why
  not?
  \begin{itemize}
  \item
    Are there specific channels they don't have access to?
  \item
    Have there any been any issues with staff or non workers in these
    channels?
  \item
    If staff are not allowed in specific channels, do they know about
    these communication channels?
  \end{itemize}
\item
  Does your union have any security concerns when using the
  communication platform? What are the perceived security risks of the
  communication channels?
\item
  (Yes - Security Risks) How does your union manage each of these
  security risks, in terms of both technical features/implementations
  and community/non-technical use implementations?
\item
  (No - Security Risks) Can you enumerate on the lack of security risks
  a bit more?
  \begin{itemize}
  \item
    Do you feel like there are no serious risks your org faces, or do
    you feel like the security is managed well enough that you
    don\textquotesingle t think it\textquotesingle s a problem?
  \end{itemize}
\item
  Is there anything else you think is important to share about the
  design and structure of the communication technology and how it
  relates to your union?
\end{itemize}

\subsection{Questions on Dynamics and
Tech}\label{questions-on-dynamics-and-tech}
\begin{itemize}
\item
  How often do you read, respond, and engage with discussions on the
  communication channels that you listed previously?
  \begin{itemize}
  \item
    Are there specific areas/activities that you are more likely to
    engage with, less likely to engage with?
  \item
    Are there specific channels that you read more or less? What is the
    reason for the difference?
  \end{itemize}
\item
  Can you recall a time that you were highly active within these digital
  communication channels and if so can you tell me about it?
  \begin{itemize}
  \item
    What was the surrounding context for the discussion?
  \item
    What was the primary goal of the discussion/communication?
  \item
    What made you want to participate in this discussion more than
    normal?
  \item
    What was the outcome of said discussion/communication?
  \end{itemize}
\item
  Has there ever been a debate/discussion in which you wanted to
  participate in but felt discouraged from doing so?
\item
  For gathering information about the unions actions, decisions and
  debates how useful would you say the communication channel is?
\item
  Can you recall a time in which you had a negative experience from
  using your union\textquotesingle s chat platforms, and if so can you
  tell me about it?
\item
  When it comes to participation, both in person and online - do you
  think having access to these forms of digital communication has
  increased or lowered participation?
  \begin{itemize}
  \item
    Could you compare and contrast in person to online participation?
  \item
    Do you think having access to the digital forms
    of participation lowers or raises in person participation?
    \end{itemize}
\item
  How about participating in a leadership capacity? That is, do you feel
  like using or having access to these digital communication channels
  has increased, decreased or not changed your ability to be a leader?
\item
  When it comes to having controversial discussions are there specific
  channels that are used?
\item
  Can you think of a time in which a digital communication channel was
  the main channel for an internal union conflict, and if so tell me
  about it?
\item
  Has there ever been an incident where private messages were shared out
  of context, either within the union or outside of the union?
\item
  Are there any rules of discussion or moderation on these platforms?
  \begin{itemize}
  \item
    (Yes - Moderation)
    \begin{itemize}
    \item
      Are you aware of how these policies were created?
    \item
      What do you think the goals of these policies are?
    \item
      How are they enforced?
    \item
      How effective have they been in achieving their goals?
    \item   
      Are there any things that you think are missing in the moderation
      of your channels?
    \item
      Are there rules that you think are unnecessary?   
    \end{itemize}
  \item
    (No - Moderation)
    \begin{itemize}
    \item
      Do you feel like there should be some rules for using the
      platform?
    \item
      Has there been any discussion around developing rules of
      engagement? 
    \end{itemize}
  \item
    Overall, when it comes to moderation do you think that the
    moderation policy or lack of a moderation policy helps or hurts the
    unions goals?
    \begin{itemize}
    \item
      In terms of participation?
    \end{itemize}
  \end{itemize}
\item
  Can you think of a crisis that occurred within your union recently?
  What did the {[}platform{]} and its associated channels look like
  during the crisis?
  \begin{itemize}
  \item
    What was the context for the crisis?
  \item
    What role did the {[}platform{]} have in the crisis?
  \item
    How did the crisis resolve and what were the long term effects of
    the crisis?
  \item
    Are there other crises in which {[}platform{]} had the opposite
    effect on the outcome?
  \end{itemize}
\item
  Are there any other dynamics within the union that your use of
  communication technologies affects?
  \begin{itemize}
  \item
    If so, what are they and could you elaborate?
  \end{itemize}
\end{itemize}

\subsection{Questions on Union Culture and
Tech}\label{questions-on-union-culture-and-tech}
\begin{itemize}
\item
  When it comes to how you communicate on these digital channels are
  there any specific informal norms - that is unwritten rules of
  communication that your union tends to follow?
  \begin{itemize}
  \item
    If you are aware, what are the origins or rationale for these norms? 
  \item  
    Are there norms that are specific to the fact that you are digital?  
  \item    
    Are there norms that are specific to being in person?    
  \item   
    What is the rationale for these distinctions? 
  \end{itemize}
\item
  Do your communication channels ever use emojis, gifs or stickers?
  \begin{itemize}
  \item
    What kind of things do they use emojis for?
  \item
    Are there any emojis that mean something unique when they're used in
    your union {[}platform{]}?
  \item
    Have you ever had an interaction with someone else using emojis that
    was memorable, for either good or bad reasons and could you describe it?
  \end{itemize}
    \begin{itemize}
        \item Are there specific values that you all share?
        \item How together/unified is the union?
        \item Is there a sense of a shared vision for the union?
    \end{itemize}
\item When it comes to your use of digital communication technology how do you think it has changed the broader cultural climate?
\item If there was one thing you could change about the culture of the union what would it be and why?
\end{itemize}

\subsubsection{Less Active/Not Comfortable Thinking About Broad Strokes}

\begin{itemize}
    \item How inclusive do you feel the organization is towards individuals who are moderately involved, like yourself?
        \begin{itemize}
            \item How do you think your union\textquotesingle s usage of digital communication channels changes that and if so how?
        \end{itemize}
    \item Do you feel there are pressures to conform in certain ways, even if you're only moderately involved?
        \begin{itemize}
            \item Where do you think the pressure to conform is coming from---leadership, peers, or the broader culture of the organization?
            \item How does that play out on the digital communication channels?
        \end{itemize}
    \item What motivates you to stay moderately involved in the organization?
\end{itemize}

\subsubsection{Leader/More Active Questions}

\begin{itemize}
    \item Can you think of a time when your union\textquotesingle s use of {[}platform{]} helped to create a sense of belonging within your union and if so can you tell me about it?
    \item Can you think of a time when your union's use of {[}platform{]} created a larger division within your union, and if so can you tell me about it?
    \item How would you say that your use of digital communication technology has changed the effectiveness of your leadership?
\end{itemize}

\subsubsection{Everyone gets asked this}

\begin{itemize}
    \item When it comes to the culture of your union and the use of communication technology, is there anything that we haven't covered today that you think is important to understanding how your union uses communication technology?
\end{itemize}

\end{document}